\documentclass[journal]{IEEEtran}
\usepackage{cite}
\usepackage{amsmath,amssymb,amsfonts}
\usepackage{graphicx}
\usepackage{textcomp}
\usepackage{multirow}%
\usepackage{mathrsfs}%
\usepackage{xcolor}%
\usepackage{manyfoot}%
\usepackage{booktabs}%
\usepackage{listings}%
\usepackage{subcaption}
\usepackage{siunitx}
\usepackage{pdflscape}

\usepackage{enumitem}
\usepackage{tikz}
\usepackage{circuitikz}
\usepackage{tabularx}
\usepackage{soul}
\usepackage{makecell}
\usetikzlibrary{positioning}
\usepackage [autostyle, english = american]{csquotes}
\MakeOuterQuote{"}
\usepackage[caption=false]{subfig}
\def\BibTeX{{\rm B\kern-.05em{\sc i\kern-.025em b}\kern-.08em
    T\kern-.1667em\lower.7ex\hbox{E}\kern-.125emX}}
\begin{document}
\title{SMURF: Scalable method for unsupervised reconstruction of flow in 4D flow MRI}
\author{Atharva Hans, Abhishek Singh, Pavlos Vlachos, and Ilias Bilionis
\thanks{This project was supported by Eli Lilly and Company, the National Science Foundation under Grant No. 2347472, and the National Institutes of Health under Award R01 HL115267. (\textit{Corresponding author: Atharva Hans.})}
\thanks{Atharva Hans, Abhishek Singh, Pavlos Vlachos, and Ilias Bilionis are with the School of Mechanical Engineering, Purdue University, West Lafayette, IN 47906, USA (e-mails: hans1@purdue.edu; sing1062@purdue.edu; pvlachos@purdue.edu; ibilion@purdue.edu).}}

\maketitle

\begin{abstract}
We introduce SMURF, a scalable and unsupervised machine learning method for simultaneously segmenting vascular geometries and reconstructing velocity fields from 4D flow MRI data. 
SMURF models geometry and velocity fields using multilayer perceptron-based functions incorporating Fourier feature embeddings and random weight factorization to accelerate convergence. 
A measurement model connects these fields to the observed image magnitude and phase data. 
Maximum likelihood estimation and subsampling enable SMURF to process high-dimensional datasets efficiently.
Evaluations on synthetic, \textit{in vitro}, and \textit{in vivo} datasets demonstrate SMURF's performance. 
On synthetic internal carotid artery aneurysm data derived from CFD, SMURF achieves a quarter-voxel segmentation accuracy across noise levels of up to 50\%, outperforming the state-of-the-art segmentation method by up to double the accuracy. 
In an \textit{in vitro} experiment on Poiseuille flow, SMURF reduces velocity reconstruction RMSE by approximately 34\% compared to raw measurements. 
In \textit{in vivo} internal carotid artery aneurysm data, SMURF attains nearly half-voxel segmentation accuracy relative to expert annotations and decreases median velocity divergence residuals by about 31\%, with a 27\% reduction in the interquartile range. 
These results indicate that SMURF is robust to noise, preserves flow structure, and identifies patient-specific morphological features. 
SMURF advances 4D flow MRI accuracy, potentially enhancing the diagnostic utility of 4D flow MRI in clinical applications.
\end{abstract}


\section{Introduction}\label{sec:introduction}

\IEEEPARstart{C}{ardiovascular} diseases lead the mortality rates globally, with conditions such as coronary artery disease, heart failure, and stroke accounting for a significant proportion of deaths each year. 
In the United States alone, heart disease causes nearly 20\% of all adult fatalities.
Additionally, congenital heart defects affect approximately 1\% of newborns, necessitating ongoing medical care throughout their lives~\cite{CHDstat1}. 
Early detection and accurate diagnosis of cardiovascular conditions are critical for improving patient outcomes and reducing mortality. 
Advanced imaging techniques, particularly 4D flow MRI~\cite{markl2014_4d}, offer valuable insights into the cardiovascular system's morphology and flow dynamics through non-invasive measurements. 
These measurements allow clinicians to comprehensively evaluate the functional and structural integrity of the anatomy of interest.
As the prevalence of cardiovascular diseases continues to rise, there is an increasing demand for reliable and efficient diagnostic tools that can deliver effective treatment planning and disease progression monitoring, thereby improving patient outcomes.

4D flow MRI advances clinical imaging by capturing time-resolved, three-dimensional velocity fields of blood flow.
From a fluid mechanics perspective, this technique enables detailed visualization and quantification of complex hemodynamic patterns, such as vortices and turbulent flows~\cite{Nayak2015_PCMRI, maroun2023_4D}. 
Beyond velocity field measurements, 4D flow MRI enables the characterization of hidden hemodynamic shear and normal stresses (pressure), inferred using conservation laws. 
These stresses are critical in evaluating functional variations in vascular physiology and blood's interaction with surrounding tissues. 
Deviations from normal hemodynamic function are markers of cardiovascular pathology, such as aneurysms, valvular heart disease, and congenital heart defects~\cite{zhuang2021role, lawley20184d}.

By capturing time-resolved data, 4D flow MRI facilitates the dynamic volumetric segmentation of vascular structures, enhancing sensitivity to anatomical abnormalities.
This segmentation process establishes vascular boundaries essential for computing wall shear stress (WSS), a biomarker that relates to endothelial behavior, vascular remodeling, and aneurysm progression~\cite{zhang2022wall, groen2008high, markl2010vivo}.
Additionally, once segmented, these structures support the analysis of intravascular pressure distributions, which estimate pressure gradients and vascular resistance~\cite{zhang2022wall}. 
Kinetic energy losses and turbulence metrics help quantify flow disturbances and assess hemodynamic efficiency.
By integrating these metrics, 4D flow MRI provides clinicians insights into vascular function, aiding pathology identification and intervention planning~\cite{stankovic20144d, garcia2019role, singh2025automated, singh2025dynamic}.

Despite clinical benefits, several challenges limit the broader adoption of 4D flow MRI. 
In the absence of contrast enhancement, poor and inhomogeneous signals require long acquisition times, causing patient discomfort and reducing clinical throughput. 
Imaging artifacts degrade data quality and complicate flow analysis.
Limited spatial resolution, often around \( 1 \, \text{mm} \) in voxel size, creates uncertainties when defining vascular boundaries and resolving velocities in complex flow regions. 
Current segmentation and velocity reconstruction methods rely heavily on manual expert annotations and isolated processes, introducing workflow inefficiencies, variability, and errors. 
These factors directly affect the accuracy of derived hemodynamic parameters.
A scalable, unsupervised method that automatically segments vascular geometries and reconstructs velocity fields at higher resolutions can address these problems.

Advancements in 4D flow MRI include methods for super-resolution and segmentation. 
Super-resolution methods like 4DFlowNet~\cite{ferdian20204dflownet} and the physics-informed approach by Fathi et al.~\cite{fathi2020super} enhance spatial resolution and reduce noise. 
4DFlowNet uses deep learning models trained on synthetic data from computational fluid dynamics simulations but may not generalize due to limited training conditions. 
Fathi et al.'s method integrates fluid dynamics principles with neural networks but requires manual tuning of physics terms, reducing adaptability. 
Both approaches focus only on resolution improvements without addressing vascular segmentation.

Unsupervised segmentation methods such as Pseudo Complex Difference (PCD) and Standardized Difference of Means (SDM) classify flow and background based on statistical distributions without training data~\cite{schnell_pcd, rothenberger2023_4DflowSDM}. 
PCD uses magnitude and phase images for threshold-based segmentation, making accuracy threshold-dependent. 
SDM uses phase images assuming zero-mean Gaussian background noise, which is effective in cerebral and aortic vasculature but sensitive to hyperparameter settings. 
Both require sufficient flow and tissue voxel counts and assume stationary vessel walls.

Supervised segmentation approaches, such as deep learning–based aortic segmentation~\cite{berhane2020fully} and MedSAM~\cite{ma2024segment}, address different segmentation challenges. 
While fully automated, the former requires large, annotated training datasets, limiting its generalizability to other anatomical regions. 
Meanwhile, MedSAM extends segmentation to multiple imaging modalities but relies heavily on expert annotations and may not capture dynamic flow complexities without additional supervision. 
Although these methods address distinct aspects of 4D flow MRI, none simultaneously offers unsupervised, high-resolution velocity reconstruction and fully automatic segmentation.

The limitations of current reconstruction and super-resolution techniques for 4D flow MRI highlight the need for an integrated approach to enhance resolution while accurately segmenting geometries and reconstructing velocity fields without depending on extensive expert inputs. 
We require an automatic, scalable, and unsupervised reconstruction method capable of effectively handling high-dimensional 4D flow MRI data to advance clinical diagnostics and streamline workflows. 

In this work, we introduce SMURF (\textbf{S}calable \textbf{M}ethod for \textbf{U}nsupervised \textbf{R}econstruction of \textbf{F}low in 4D flow MRI), a method designed to automatically and simultaneously segment vascular geometries and reconstruct velocity fields from 4D flow MRI data, with the ability to super-resolve both the segmented geometry and the reconstructed velocity field. 
By representing geometry and velocity fields through functions parameterized by a modified multilayer perceptron (MLP) architecture~\cite{wang2021understanding}, we achieve continuous representations that can be evaluated at arbitrarily fine resolutions, enabling super-resolution of the segmented geometries and reconstructed velocity fields.
To better capture the complexities of 4D flow MRI data and ensure stable convergence, we employ techniques such as Fourier feature embedding~\cite{tancik2020fourier} and random weight factorization~\cite{wang2022random} within the MLPs.
We then define a measurement model that intertwines the geometry and velocity fields, mapping these representations to the observed 4D flow MRI measurements. 
This integrated measurement model offers four specific advantages: (i) by jointly modeling geometry and velocity, we can perform segmentation and velocity reconstruction simultaneously in a single step, improving efficiency and coherence; (ii) the method utilizes both magnitude and velocity data from 4D flow MRI, resulting in more accurate segmentation and velocity reconstruction by leveraging complementary information; (iii) we can operate using either both magnitude and velocity data or solely velocity data, enhancing applicability across various clinical scenarios and conditions regarding data availability; and (iv) our approach is unsupervised, eliminating the need for expert annotations, streamlining the 4D flow processing workflow, and reducing potential sources of error.
We employ maximum likelihood estimation and subsampling techniques to process high-dimensional image data efficiently.

We structure the paper as follows.
We outline the mathematical details of our methodology in Section~\ref{sec:method}, describe the datasets used for validation, present our results in Section~\ref{sec:results}, and conclude in Section~\ref{sec:conclusion}.

\section{Method}\label{sec:method}
At each time step \(t\), taking values from 1 to \(N_t\), and for each slice \(s\), taking values from 1 to \(N_s\), within the MRI scanner, 4D flow MRI data consist of one magnitude image and three phase images, each comprising \(N_r\times N_c\) pixels. 
This results in magnitude and phase data across \(N_v=N_s N_r N_c\) voxels at every time step. 
We denote the center of voxel \(i\), taking values from 1 to \(N_v\), as \(\mathbf{r}_i = (r_{i1}, r_{i2}, r_{i3})\), with \(\mathbf{R}=(\mathbf{r}_1, \mathbf{r}_2, \ldots, \mathbf{r}_{N_v})\) representing all voxel centers.

4D flow MRI utilizes phase contrast imaging, wherein the recorded phase values linearly correlate with biofluid flow velocities when phase wrapping artifacts are absent. 
We convert phase image values to velocity using the velocity encoding (venc) parameter that the scanner sets, according to the equation: $\text{velocity} = \frac{\text{phase} \times \text{venc}}{\pi}.$
Herein, we work directly with the velocity values.

We organize all magnitude data into a \(N_v\times N_t\) matrix \(\mathbf{Y}^{\text{mag}}\), with elements \(y^{\text{mag}}_{it}\) indicating the magnitude at voxel \(i\) at time step \(t\). 
We arrange all velocity data in a \(N_v\times N_t\times 3\) tensor \(\mathbf{Y}^{\text{vel}}\), where \(y^{\text{vel}}_{itk}\) represents the \(k\)-th velocity component (with \(k\) taking values from 1 to 3) at voxel \(i\) at time step \(t\).

\subsection{Class representation}
We operate within the 3D domain \( \Omega \subset \mathbb{R}^3 \), which we model as consisting of \( M \) distinct classes.
Each class corresponds to a region with unique physical properties and imaging features, such as signal intensity and velocity behavior.
Explicitly defining these classes allows our method to leverage differences in the statistical distributions of magnitude and velocity signals across various tissue types and fluid regions.

We represent these classes using \( m \), where \( m \) ranges from 1 to \( M \) and identifies the \( m \)-th class.
We focus on the time interval \( [0, T] \) and represent the fields within an MRI voxel index.
The \( M \)-dimensional function \( \mathbf{g}_{\psi}: \Omega \times [0, T] \to \mathbb{R}^{M} \), parameterized by \( \psi \), captures the geometry of these classes.
The function \( \mathbf{g}_{\psi}(\mathbf{r}_i, t) \) generates values that, through a sparsemax function~\cite{martins2016softmax}, determine the probabilities of voxel \( i \) at time step \( t \) belonging to each class.

We define \( z_{it} \) as a latent discrete variable for the class type at voxel \( i \) at time step \( t \). 
The distribution of \( z_{it} \), modeled as categorical with probabilities for each class type \( m \), is determined by applying sparsemax to the outputs of \( \mathbf{g}_{\psi} \):
\begin{equation}
p\left( z_{it} = m \mid \psi \right) = \text{sparsemax}\left( \mathbf{g}_{\psi}(\mathbf{r}_i, t) \right)_m.
\label{eq:material_probability}
\end{equation}
The sparsemax function promotes sparsity in the output probabilities, effectively assigning zero probability to unnecessary classes. 
This property helps in automatically selecting the number of classes from \( M \) during optimization.

We represent \( \mathbf{g}_{\psi} \) using a modified MLP~\cite{wang2021understanding} with Fourier feature embeddings~\cite{tancik2020fourier} and random weight factorization (RWF)~\cite{wang2022random}. We provide the details of the network structure in Appendix~\ref{sec:neural_networks_archs}.

To perform segmentation, we use the geometry field \( \mathbf{g}_{\psi} \). 
We determine the probability of each voxel belonging to a specific class using \eqref{eq:material_probability} and assign each voxel to the class with the highest probability. 
Extracting the voxels belonging to the same class enables segmentation.

Unlike traditional segmentation methods that produce binary masks (0 for tissue, 1 for flow), our framework offers probabilities across all classes for each voxel, reflecting uncertainty in class assignments. 
This probabilistic representation naturally accommodates partial volume effects, where a voxel contains signals from multiple tissue or fluid types.
While this work focuses on selecting the most probable class per voxel to evaluate against traditional segmentation methods, future research will investigate the probabilistic capabilities to enhance interpretation and support more informed clinical decision-making.

Next, we connect the latent variable to the MRI data via a likelihood function.

\subsection{MRI magnitude data likelihood}
For the MR image magnitude modality, we model the data using a Gaussian likelihood, where both mean and variance depend on the latent variable representing the class type. 
This allows us to reflect each class type's unique properties accurately. 
We express this likelihood as:
\begin{equation}
    p\left(y^{\text{mag}}_{it} \mid z_{it}=m\right) = \mathcal{N}\left(y^{\text{mag}}_{it} \mid \mu_{m}, \sigma^2_{m}\right).
    \label{eq:mag_material_probability}
\end{equation}
Here, \(\mathcal{N}(y^{\text{mag}}_{it} \mid \mu_m, \sigma^2_m)\) denotes the Gaussian probability density function evaluated at \(y^{\text{mag}}_{it}\), with mean \(\mu_m\) and variance \(\sigma^2_m\).
We treat \(\mu_m\) and \(\sigma^2_m\) as learnable parameters and infer them from the data.
There are \(M\) mean and \(M\) variance parameters for the \(M\) different classes.

Applying the sum rule of probability, along with \eqref{eq:material_probability} and \eqref{eq:mag_material_probability}, establishes the marginal conditional probability of $y^{\text{mag}}_{it}$ based on $\psi$:
\begin{equation}
p\left(y^{\text{mag}}_{it} \mid \psi\right) = \sum_{m=1}^{M} \left[ p\left(y^{\text{mag}}_{it} \mid z_{it} = m\right) \cdot p\left(z_{it} = m \mid \psi\right) \right].
\label{eq:mag_data_probability}
\end{equation}
Assuming conditional independence of the measurement noise, the likelihood for the magnitude data becomes:
\begin{equation}
\begin{aligned}
p(\mathbf{Y}^{\text{mag}} \mid \psi) = \prod_{t=1}^{N_t} \prod_{i=1}^{N_v} p\left(y^{\text{mag}}_{it} \mid \psi\right).
\end{aligned}
\label{eq:all_mag_data_probability}
\end{equation}

\subsection{MRI velocity data likelihood}
We model velocity as a three-dimensional function 
\(\mathbf{v}_{\boldsymbol{\phi}}: \Omega \times [0, T] \to \mathbb{R}^3\), 
where \(\boldsymbol{\phi} = (\phi_1, \phi_2, \phi_3)\) collects the distinct parameter sets for the three velocity components. 
Hence, \(\mathbf{v}_{\boldsymbol{\phi},k}(\mathbf{r}_i, t)\) denotes the \(k\)-th component at voxel \(i\) and time step \(t\). 
We employ three separate modified MLPs—one for each velocity component—each with its parameters \(\phi_k\) (\(k=1,2,3\)) and equipped with Fourier feature embeddings and RWF. 
This approach yields improved performance compared to using a single neural network to jointly model all three velocity components.
Appendix~\ref{sec:neural_networks_archs} provides the detailed model architectures.

We use a Gaussian likelihood for modeling the velocity data:
\begin{equation}
p\left(\mathbf{y}^{\text{vel}}_{it} \mid z_{it}=m, \boldsymbol{\phi}\right) = \mathcal{N}\left( \mathbf{y}^{\text{vel}}_{it} \mid \mathbf{v}_{\boldsymbol{\phi}}(\mathbf{r}_i, t), \mathbf{\Sigma}_m \right).
\label{eq:vel_material_probability}
\end{equation}
Here \(\mathbf{y}^{\text{vel}}_{it} = (y^{\text{vel}}_{it1}, y^{\text{vel}}_{it2}, y^{\text{vel}}_{it3})\) represents the velocity vector at voxel \(i\) and time \(t\) and \( \mathbf{\Sigma}_m \) is a diagonal covariance matrix representing measurement noise for class $m$.

Incorporating \eqref{eq:material_probability} and \eqref{eq:vel_material_probability} through the sum rule of probability, the marginal conditional probability of $\mathbf{y}^{\text{vel}}_{it}$ based on $\psi$ and $\boldsymbol{\phi}$ becomes:
\begin{equation}
p\left(\mathbf{y}^{\text{vel}}_{it} \mid \psi, \boldsymbol{\phi}\right) = \sum_{m=1}^{M} \left[ p\left(\mathbf{y}^{\text{vel}}_{it} \mid z_{it} = m, \boldsymbol{\phi} \right) \cdot p\left(z_{it} = m \mid \psi\right) \right].
\label{eq:vel_data_probability}
\end{equation}
Assuming conditional independence across the velocity data, the likelihood for the velocity data becomes:
\begin{equation}
\begin{aligned}
p(\mathbf{Y}^{\text{vel}} \mid \psi, \boldsymbol{\phi}) = \prod_{t=1}^{N_t} \prod_{i=1}^{N_v} p\left(\mathbf{y}^{\text{vel}}_{it} \mid \psi, \boldsymbol{\phi}\right).
\end{aligned}
\label{eq:all_vel_data_probability}
\end{equation}

\subsection{Parameter estimation}\label{sec:param_estimation}
Our method consists of field parameters \{$\boldsymbol{\phi}$, $\psi$\}, likelihood parameters for magnitude data \(\{\mu_m, \sigma^2_m\}_{m=1}^M\), and likelihood parameters for velocity data, which include diagonal covariance matrices \(\{\mathbf{\Sigma}_m\}_{m=1}^{M}\).
Collectively, we denote these parameters using \(\theta = \{\psi, \boldsymbol{\phi}, (\mu_m, \sigma^2_m)_{m=1}^M, \{\mathbf{\Sigma}_m\}_{m=1}^{M}\}.\)

We determine the optimal values of \(\theta\) through maximum likelihood estimation.
To ensure numerical stability, we operate on log probabilities. 
The loss function combines negative log-likelihoods for both magnitude and velocity data:
\begin{equation}
\begin{split}
    \text{Loss}_{\text{data}}(\theta, \mathbf{Y}^{\text{mag}}, \mathbf{Y}^{\text{vel}}) = & -\log p(\mathbf{Y}^{\text{mag}} \mid \psi) \\
    & - \log p(\mathbf{Y}^{\text{vel}} \mid \psi, \boldsymbol{\phi}).
\end{split}
\label{eq:loss_function_data}
\end{equation}
Using \eqref{eq:all_mag_data_probability} and \eqref{eq:all_vel_data_probability}, we write \eqref{eq:loss_function_data} as:
\begin{equation}
\begin{split}
    \text{Loss}_{\text{data}}(\theta, \mathbf{Y}^{\text{mag}}, \mathbf{Y}^{\text{vel}}) = & -\sum_{t=1}^{N_t} \sum_{i=1}^{N_v} \log p\left(y^{\text{mag}}_{it} \mid \psi\right) \\
    & - \sum_{t=1}^{N_t} \sum_{i=1}^{N_v} \log p\left(\textbf{y}^{\text{vel}}_{it} \mid \psi, \boldsymbol{\phi}\right).
\end{split}
\label{eq:loss_function_data_simplified}
\end{equation}

We also introduce an \(L_2\) regularization for the neural network parameters. 
This regularization is given by:
\begin{equation}
    \text{Loss}_{l_2}(\psi, \boldsymbol{\phi}) = \|\psi\|^2_2 + \|\boldsymbol{\phi}\|^2_2,
    \label{eq:loss_function_l2}
\end{equation}
where \( \| \cdot \|_2 \) represents the \( L_2 \) norm.
Incorporating \(L_2\) regularization promotes simpler models by penalizing larger parameter values, thus preventing overfitting.

Minimizing the total loss:
\begin{equation}
\begin{aligned}
    \text{Loss}_{\text{total}}(\theta, \mathbf{Y}^{\text{mag}},
    \mathbf{Y}^{\text{vel}}) = \text{Loss}_{\text{data}}(\theta, \mathbf{Y}^{\text{mag}}, \mathbf{Y}^{\text{vel}}) \\
    + \alpha_1 \, \text{Loss}_{l_2}(\psi, \boldsymbol{\phi}),
\end{aligned}
\label{eq:loss_function_total}
\end{equation}
allows for simultaneous segmentation of geometry and reconstruction of velocity fields while enabling super-resolution of both.
Here, \(\alpha_1\) is the weight for the regularization term.

\subsubsection{Velocity data-driven reconstruction}
Our method can incorporate MRI magnitude and velocity data, as shown in \eqref{eq:loss_function_data}. 
However, magnitude data are not necessary due to our measurement model's design. 
The intertwined nature of the geometry and velocity field allows for reconstructions solely based on velocity data, increasing flexibility in applications where magnitude data is unavailable or unreliable. 
The modified total loss function for velocity data-driven reconstructions becomes:
\(
\text{Loss}^{\text{vel}}_{\text{total}}(\theta^{\text{vel}}, \mathbf{Y}^{\text{mag}}, \mathbf{Y}^{\text{vel}}) = - \sum_{t=1}^{N_t} \sum_{i=1}^{N_v} \log p\bigl(\mathbf{y}^{\text{vel}}_{it} \mid \psi, \boldsymbol{\phi}\bigr) + \alpha_1\, \text{Loss}_{l_2}(\psi, \boldsymbol{\phi})
\),
where \(\theta^{\text{vel}} = \{\psi, \boldsymbol{\phi}, \{\mathbf{\Sigma}_m\}_{m=1}^{M}\}\).

\subsubsection{Magnitude data-driven segmentation}
Our method also allows for segmentation using only magnitude data, demonstrating its adaptability. 
This functionality is especially relevant when velocity data does not significantly impact the segmentation outcome.
The total loss function for magnitude data-driven segmentation adjusts as follows: 
\(
\text{Loss}^{\text{mag}}_{\text{total}}(\theta^{\text{mag}}, \mathbf{Y}^{\text{mag}}) = - \sum_{t=1}^{N_t} \sum_{i=1}^{N_v} \log p\left(y^{\text{mag}}_{it} \mid \psi\right) + \alpha_1 (\|\psi\|^2_2)
\),
where \(\theta^{\text{mag}} = \{\psi, \{\mu_m, \sigma^2_m\}_{m=1}^M\}\).

\subsection{Scaling to large datasets}\label{sec:scaling_large_datasets}
Evaluating the loss in~\eqref{eq:loss_function_data} scales poorly with dataset size due to the complexity of computing log densities~\cite{hans2020quantifying, hans2023bayesian}. 
To mitigate this, we employ subsampling~\cite{hans2023stochastic, hans2024bayesian}, which exploits the conditional independence of magnitude and velocity values given $\psi$ and $\boldsymbol{\phi}$.

Subsampling enables us to approximate log-likelihood terms in the loss function as follows: 
\(
\log p(\mathbf{Y}^{\text{mag}} \mid \psi) \approx \frac{N_t N_v}{S_t S_v}\sum_{t \in \mathcal{I}_{t}} \sum_{i \in \mathcal{I}_{v}} \log p\left(\mathbf{Y}^{\text{mag}}_{it} \mid \psi\right)
\)
and 
\(
\log p(\mathbf{Y}^{\text{vel}} \mid \psi, \boldsymbol{\phi}) \approx \frac{N_t N_v}{S_t S_v}\sum_{t \in \mathcal{I}_{t}} \sum_{i \in \mathcal{I}_{v}} \log p\left(\mathbf{Y}^{\text{vel}}_{it} \mid \psi, \boldsymbol{\phi}\right)
\),
where \(\mathcal{I}_{t}\) and \(\mathcal{I}_{v}\) represent randomly selected mini-batches of indices, sized \(S_t\) (where \(S_t < N_t\)) and \(S_v\) (where \(S_v < N_v\)), respectively. Note that by keeping \(S_t\) and \(S_v\) fixed, the computational cost for each optimization step will be independent of the dataset size.

\subsection{Optimizer and parameter initialization}
We implement our loss function using the \texttt{JAX} library~\cite{jax2018github} to leverage its automatic differentiation capabilities.
We develop our neural network architectures using the \texttt{Equinox}~\cite{kidger2021equinox} library, which ensures full compatibility with \texttt{JAX}. 
For optimization, we employ the ADAM optimizer~\cite{kingma2014adam} from the \texttt{Optax} library, with $\beta_1=0.9$, $\beta_2=0.999$, $\epsilon=10^{-8}$, and a constant learning rate of $10^{-3}$.

Before optimization, we perform the following steps:
\begin{itemize}
    \item Standardize observed data to have a zero mean and unit variance.
    \item Set the number of classes \(M\) to six for both synthetic and experimental datasets. Using sparsemax effectively prunes any surplus classes, ensuring the results are invariant to an \(M\) larger than the true number of classes.
    \item Initialize the magnitude mean parameters \(\{\mu_m\}_{m=1}^M\) to be evenly spaced between \(-1.0\) and \(1.0\). Set the variance parameters \(\{\sigma^2_m\}_{m=1}^M\) and the diagonal elements of the velocity covariance matrices \(\{\mathbf{\Sigma}_m\}_{m=1}^{M}\) to 1.0. The method remains robust to these initial values as long as the data is standardized.
    \item Choose subsample sizes \(S_t\) and \(S_v\) as integer values, representing 10\% of the total number of time steps (\(N_t\)) and voxels (\(N_v\)), respectively.
    \item Set the regularization weight \(\alpha_1\) to 250, based on empirical tuning to best recover the true geometry and velocity field in noisy synthetic data, as measured by \(L^2\)-error. 
\end{itemize}

We optimize over 250 epochs for the synthetic dataset and 500 epochs for the \textit{in vitro} and the \textit{in vivo} datasets.
Using one NVIDIA A100 GPU with 40 GB of memory, the optimization takes approximately 0.9 minutes for the synthetic case and 1.2 minutes for the \textit{in vitro} and the \textit{in vivo} cases.

Further details regarding neural network parameter initialization appear in Appendix~\ref{sec:neural_networks_archs}.

\section{Results}\label{sec:results}
\subsection{Datasets}
\subsubsection{Synthetic internal carotid artery aneurysm flow}
We generate synthetic 4D flow MRI data using CFD simulations of an internal carotid artery (ICA) aneurysm as ground truth. For a detailed description of the simulation setup, refer to \cite{brindise_multimodality}.

To simulate intravoxel averaging effects caused by finite k-space sampling, we convolve the CFD velocity field with a 3D sinc kernel, representing the k-space point-spread function \cite{rispoli_synMRI}. 
For 4D flow voxel dimensions $\Delta x$, $\Delta y$, $\Delta z$, we define the kernel as: 
$K(x,y,z)=\operatorname{sinc}(x/\Delta x)\operatorname{sinc}(y/\Delta y)\operatorname{sinc}(z/\Delta z)$, 
where $\operatorname{sinc}(u)=\frac{\sin(\pi u)}{\pi u}$. 
We set $K(x,y,z)=0$ if $|x|>2\Delta x$, $|y|>2\Delta y$, or $|z|>2\Delta z$.

We apply this convolution to individual signal components \( s \), representing the signal magnitude and velocities, mapping them from the CFD grid to the 4D flow grid: \( s_{\text{blurred}}(\mathbf{r}_{i}) = \left[ s(\mathbf{r}_{\text{CFD}}) * K \right](\mathbf{r}_{i}) \), where \( \mathbf{r}_{i} \) denotes the center of the \( i \)-th voxel in the 4D flow grid, and \( \mathbf{r}_{\text{CFD}} \) represents the centers of the CFD grid points.

Since the CFD data lacks a signal magnitude field, we assign a value of 1 in the flow regions and 0.1 in the background regions to create a 10:1 contrast ratio.  

We generate the synthetic 4D flow MRI data as: \( s_{\text{4DFlow}}(\mathbf{r}_{i}) = s_{\text{blurred}}(\mathbf{r}_{i}) + \epsilon_i \), where \( \epsilon_i \sim \mathcal{N}(0, \sigma) \).
We define \( \sigma \) as: \( \sigma = \alpha \max(s_{\text{blurred}}) \), with \( \alpha \) ranging from 5\% to 60\% to simulate varying noise levels.

This procedure produces 4D flow MRI data with velocity and magnitude fields that reflect in vivo noise levels and resolution constraints.

\subsubsection{In vitro Poiseuille flow}
We acquired 4D flow MRI data at Purdue MRI Facility on a GE Discovery MR750 3T scanner. 
The setup included steady Poiseuille flow at about \(30 \, \text{cm/s}\) mean velocity through a Polydimethylsiloxane (PDMS) channel of a quarter-inch diameter. 
A Compuflow 1000 MR piston pump drove the flow. 
We used a 60/40 water-glycerol mixture with a density of \(1110 \, \text{kg/m}^3\) and a viscosity of \(0.0035 \, \text{Pa}\cdot\text{s}\).

We acquired data at \(1.0 \times 1.0 \times 1.0 \, \text{mm}^3\) and apply ZIP \(\times\) 2 interpolation to \(0.5 \times 0.5 \times 0.5 \, \text{mm}^3\). 
We captured nine timepoints over the cardiac cycle, with a repetition time TR = \(9.15 \, \text{ms}\), an echo time TE = \(3.50 \, \text{ms}\), flip angle \(=8^\circ\), venc = \(70 \, \text{cm/s}\), and total scan time of about 4 minutes.

We also performed a TOF scan with \(0.5078 \times 0.5078 \times 1.0 \, \text{mm}^3\) resolution, flip angle \(=8^\circ\), TR = \(3.552 \, \text{ms}\), and TE = \(1.252 \, \text{ms}\). 
We segmented the images semi-automatically using ITK-Snap \cite{itksnap}. 
For comparisons with the 4D flow-segmented geometry, we automatically registered two surfaces using point-cloud registration from Python’s \texttt{Open3D} library~\cite{open3d}.

\subsubsection{In vivo internal carotid artery aneurysm}\label{sec:invivo_data}
An ICA aneurysm was imaged at Northwestern Memorial Hospital using a 3 T MRI scanner (Skyra, Siemens Healthcare, Erlangen, Germany). 
The protocol included TOF angiography at \(0.4 \times 0.4 \times 0.6 \, \text{mm}^3\) and 4D flow MRI with TE/TR = \(2.997/6.4 \, \text{ms}\), flip angle = \(15^\circ\), venc = \(80 \, \text{cm/s}\), temporal resolution = \(44.8 \, \text{ms}\), and spatial resolution = \(1.09 \times 1.09 \times 1.30 \, \text{mm}^3\). For further details on the processing, please refer to \cite{brindise_multimodality}.

\subsection{Verification on the synthetic 4D flow}

\begin{figure}
        \centering
        \includegraphics[width=1\linewidth]{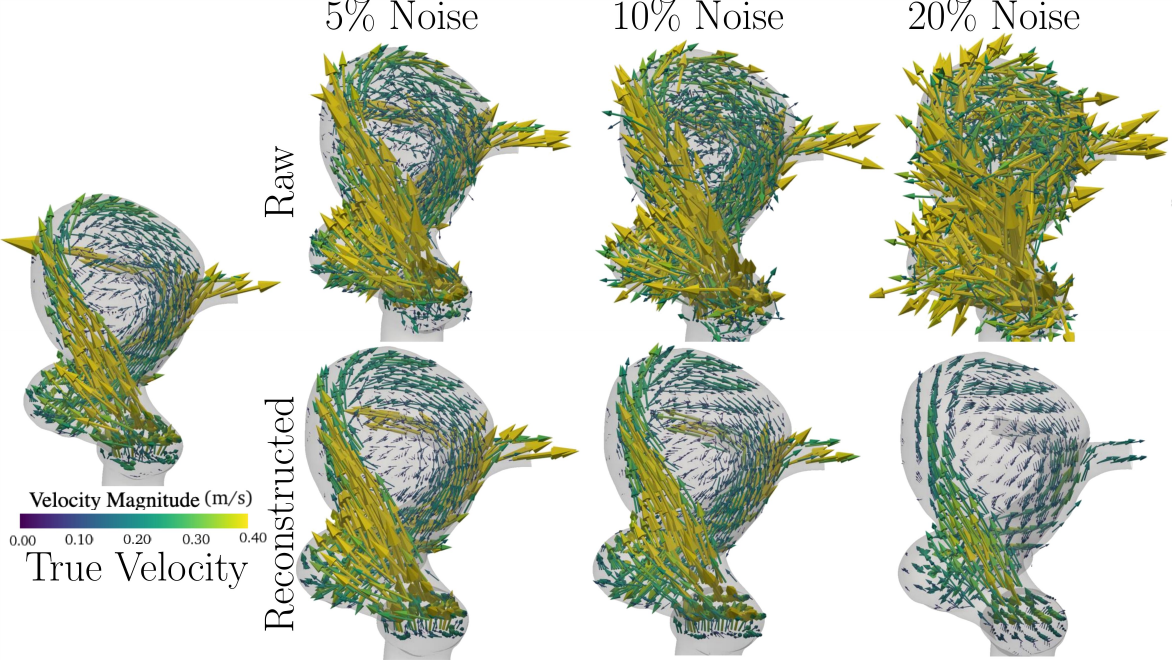}
        \caption{Visualization of velocity fields at peak systole for the synthetic ICA aneurysm dataset, showing true velocities (left column), measured velocities (top row), and SMURF-reconstructed velocities (bottom row), all overlaid on the true geometry.}
        \label{fig:syn_fig_4}
\end{figure}

We verify our method using synthetic 4D flow MRI data generated from CFD simulations of an ICA aneurysm. 
The flow in the ICA aneurysm, which contains a dome near the bifurcation, starts with jets entering the dome. 
These jets create vortices and regions of low velocity along the superior wall. The flow exits the aneurysm and branches into the posterior communicating artery, middle cerebral artery (MCA), and anterior cerebral artery (ACA). 
Vessel narrowing increases velocity in these branches (Fig.~\ref{fig:syn_fig_4}). These patterns vary over time, with peak flow during systole and reduced flow during diastole.

\begin{figure}
        \centering
        \includegraphics[width=1\linewidth]{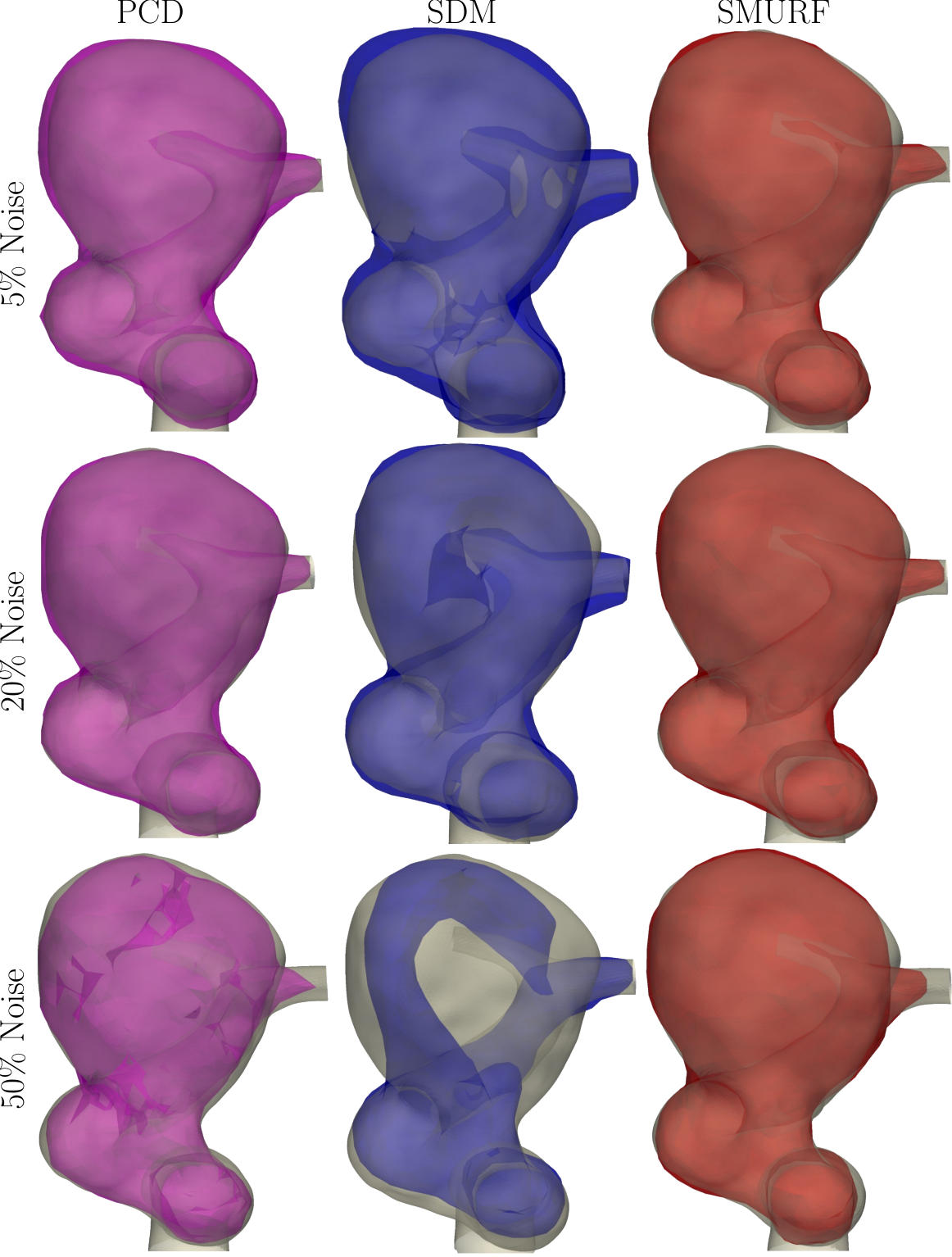}
        \caption{Segmentation comparison for the synthetic ICA aneurysm 4D flow model at increasing noise levels (5\%, 20\%, 50\%), showing ground truth (gray) with overlaid segmentations from PCD (magenta), SDM (blue), and SMURF (red).}
        \label{fig:syn_fig_1}
\end{figure}

\textbf{Flow segmentation.}
The CFD geometry is the true geometry (gray region in Fig.~\ref{fig:syn_fig_1}).
We evaluate SMURF's segmentation and velocity reconstruction under noise levels from $5\%$ to $60\%$, representing low to extreme noise conditions. 
We compare SMURF's segmentation performance with that of state-of-the-art methods—PCD and SDM, which both produce static masks. 
SMURF produces time-resolved segmentations. 

Cerebral aneurysms do not deform much during the cardiac cycle, so we construct a static SMURF mask using majority logic, classifying a voxel as flow if it appears in more than half of the time points.
This choice has a negligible effect on SMURF’s segmentation, with the average fractional change in segmented volume across cardiac phases relative to the static mask remaining below 5\% for noise levels up to 50\%.
At higher noise levels, segmentation variability increases due to greater measurement uncertainty.

We extract wall surfaces from flow-labeled 4D flow voxels by applying a Gaussian filter to the binary classification mask, then using the marching cubes algorithm on the $level = 0.5$ isocontour, and finally smoothing with the Taubin filter (50 iterations, relaxation factor 0.01) via the \texttt{Open3D}~\cite{open3d} library.

Fig.~\ref{fig:syn_fig_1} compares segmentations at noise levels of 5\%, 20\%, and 50\%.
At low noise, all methods classify flow accurately.
PCD and SDM assign partial-volume voxels to flow, leading to over-segmentation, while SMURF assigns these voxels to a separate class, improving accuracy.
SDM is the most affected by noise, misclassifying slow-flow regions such as vortex cores and boundary layers as background at higher noise levels. 
While PCD remains accurate, it struggles at $50\%$ noise, leading to a noisier segmented surface. 
Overall, SMURF produces more robust and smoother surfaces.
These improvements in segmentation accuracy and surface smoothness are critical for precise wall delineation and essential for evaluating boundary layer dynamics such as WSS. 
Future work will further investigate this capability.

\begin{figure}
        \centering
        \includegraphics[width=1\linewidth]{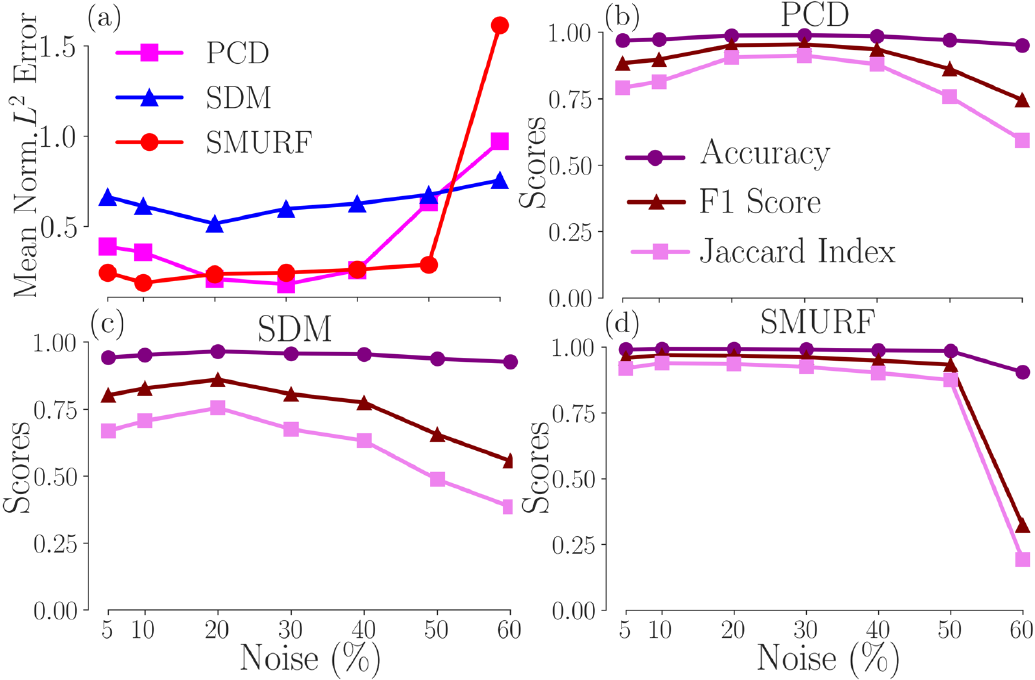}
        \caption{Quantitative comparison of segmentation accuracy for PCD, SDM, and SMURF methods across varying noise levels. 
        (a) Mean normalized \(L^2\)-error; 
        (b–d) Segmentation performance evaluated using Accuracy, F1 Score, and Jaccard Index (see Table~\ref{tab:seg_scores}).
        The X-axis for the first row of plots matches that of the second row.}
        \label{fig:syn_fig_2}
\end{figure}

Fig.~\ref{fig:syn_fig_2}a shows quantitative segmentation comparisons using mean normalized \(L^2\)-error. 
This metric measures how closely the segmented surface matches the true geometry by calculating the shortest Euclidean distance from each point on the segmented surface to the closest point on the true geometry. 
Normalizing these distances by the smallest voxel dimension ensures scale independence. 
The final error is the average of these normalized distances across all points on the segmented surface.
SMURF maintains an average error within a quarter of a voxel across a broad noise range. 
PCD closely follows SMURF’s performance, surpassing it at intermediate noise levels and 60\% noise. 
At 60\% noise, SMURF's multiclass classification loses effectiveness, likely because random fluctuations overwhelm subtle velocity signals. This causes voxels with slower flow to be misclassified relative to less noise-affected high-velocity regions, ultimately reducing segmentation accuracy.
SDM, optimized primarily for in vivo 4D flow, shows larger errors overall. 
Notably, all methods achieve sub-voxel accuracy for most of the tested range.

We further assess segmentation quality using Accuracy, F1 Score, and Jaccard Index defined in Table~\ref{tab:seg_scores}.
Fig.~\ref{fig:syn_fig_2}(b–d) shows that all methods classify background better than flow, as indicated by $\text{Accuracy} > \text{F1 Score} > \text{Jaccard Index}$. 
Interestingly, all methods achieve their best performance at 20\% noise, likely due to an interplay between the spatial blurring inherent to 4D flow (which tends to overpredict flow voxels) and the erosive effects of noise that offset this overprediction.
SMURF maintains stable performance up to 50\% noise, consistently achieving scores above 90\% across all metrics, with the ranking trend remaining $\text{SMURF} > \text{PCD} > \text{SDM}$. 
At 60\% noise, SMURF underpredicts flow, resulting in a decline in performance.
\begin{table}[ht]
    \centering
    \caption{Definitions of segmentation metrics used for quantitative evaluation.}
    \label{tab:seg_scores}
    \renewcommand{\arraystretch}{1.3}
    \small
    \begin{tabular*}{\linewidth}{@{\extracolsep{\fill}}cc}
        \toprule
        \multicolumn{1}{c}{\textbf{Metric}} & \multicolumn{1}{c}{\textbf{Definition}} \\
        \midrule
        Accuracy    & $\frac{|T \cap P| + |\neg T \cap \neg P|}{|T \cup P| + |\neg T \cup \neg P|}$ \\
        Precision   & $\frac{|T \cap P|}{|P|}$ \\
        Recall      & $\frac{|T \cap P|}{|T|}$ \\
        F1 score    & $\frac{2|T \cap P|}{2|T \cap P| + |T \setminus P| + |P \setminus T|}$ \\
        Dice score  & $\frac{2|T \cap P|}{|T| + |P|}$ \\
        Jaccard index & $\frac{|T \cap P|}{|T \cup P|}$ \\
        \midrule
        \multicolumn{2}{c}{\textbf{Symbol Definitions:}} \\
        \multicolumn{2}{c}{
          \begin{minipage}{0.8\linewidth}
            \centering
            \small
            \begin{itemize}[leftmargin=*]
              \item $T$: Voxel set classified as flow in ground truth.
              \item $P$: Voxel set classified as flow in predicted segmentation.
              \item $\neg T$: Voxel set classified as background in ground truth.
              \item $\neg P$: Voxel set classified as background in predicted segmentation.
              \item $\cap$: Intersection of two sets.
              \item $\cup$: Union of two sets.
              \item $\setminus$: Set difference.
              \item $|\cdot|$: Cardinality (number of voxels in a set).
            \end{itemize}
          \end{minipage}
        }\\
        \bottomrule
    \end{tabular*}
\end{table}

\textbf{Flow reconstruction.}
Fig.~\ref{fig:syn_fig_4} visualizes SMURF's velocity reconstruction alongside the raw and ground truth velocities at peak systole. We linearly interpolate CFD velocities, defined on a tetrahedral grid, onto the 4D flow grid for voxel-wise comparisons.  
SMURF effectively preserves key flow features, including jets entering the aneurysm and vortices, while reducing noise. 
However, as noise levels increase and signal strength diminishes, SMURF tends to underpredict velocities.

\begin{figure}
    \centering
    \includegraphics[width=1\linewidth]{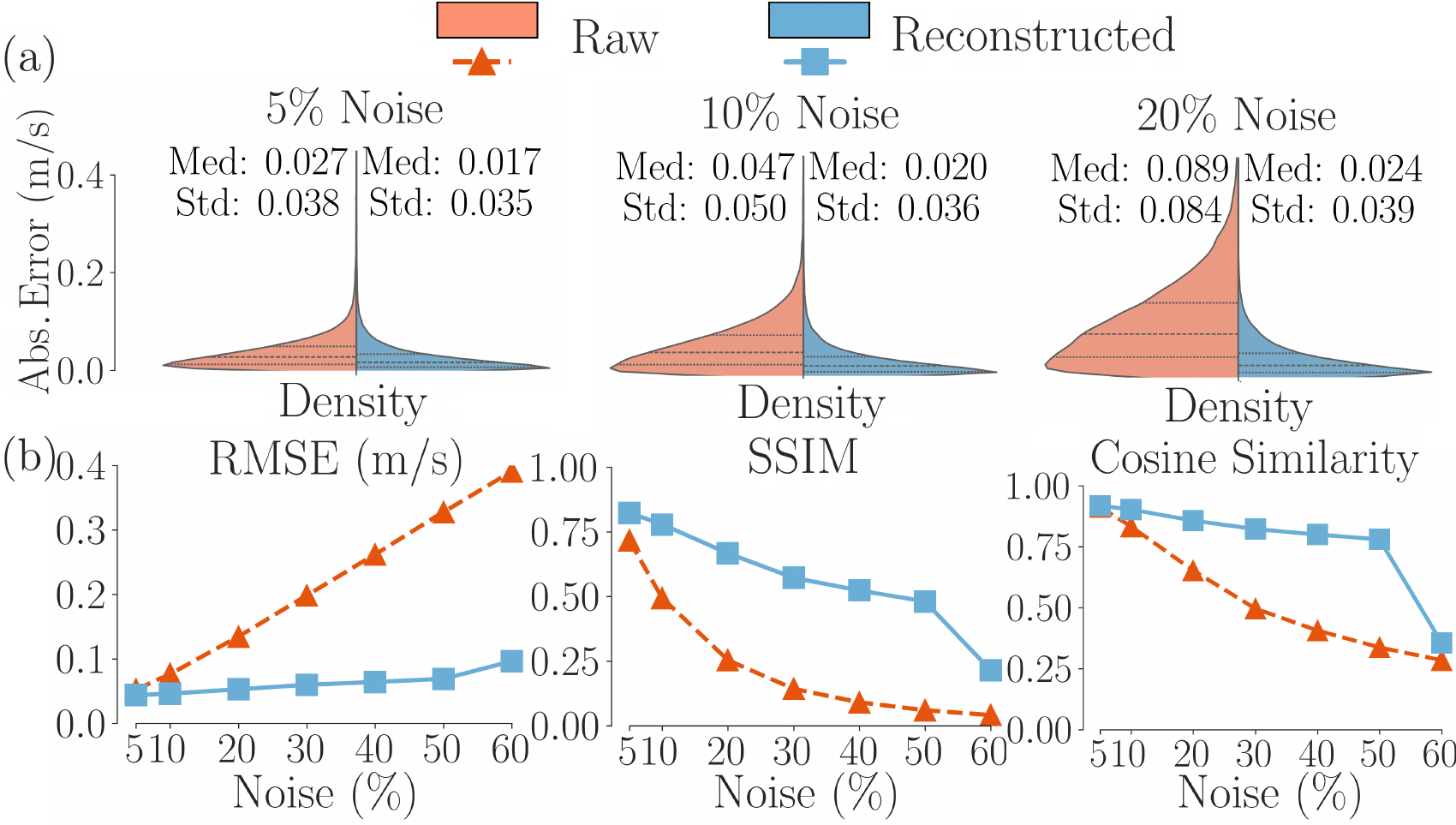}
    \caption{Quantitative evaluation of SMURF's velocity reconstruction for the synthetic ICA aneurysm dataset. 
    (a) Velocity magnitude errors for raw and reconstructed fields across 5\%, 10\%, and 20\% noise levels; 
    (b) Comparison of raw and reconstructed velocity fields using RMSE, SSIM, and cosine similarity metrics across different noise levels.}
    \label{fig:syn_fig_5}
\end{figure}

Fig.~\ref{fig:syn_fig_5} quantitatively assesses SMURF's reconstruction performance. Violin plots in Fig.~\ref{fig:syn_fig_5}a compare velocity magnitude errors for raw and reconstructed velocity fields across varying noise levels. Raw velocities exhibit a nearly linear increase in noise-induced errors, with a $229.6\%$ increase in median error from $5\%$ to $20\%$ noise. In contrast, SMURF effectively mitigates noise, demonstrating only a $41.1\%$ increase in median error, resulting in more stable and accurate velocity reconstructions.

Fig.~\ref{fig:syn_fig_5}b evaluates reconstruction performance using root mean squared error (RMSE), structural similarity index (SSIM), and cosine similarity for velocity magnitudes relative to the true velocities.
SMURF reduces RMSE up to 4 times over the tested range, demonstrating improved accuracy. 
Up to 50\% noise, SMURF substantially restores SSIM from near-zero values in high-noise cases, reflecting strong structural similarity with the true flow fields. 
Cosine similarity scores further highlight SMURF's ability to recover flow directionality, maintaining values above 80\% up to 50\% noise.

%
\subsection{Validation on the in vitro Poiseuille flow} 
\begin{figure}
        \centering
        \includegraphics[width=1\linewidth]{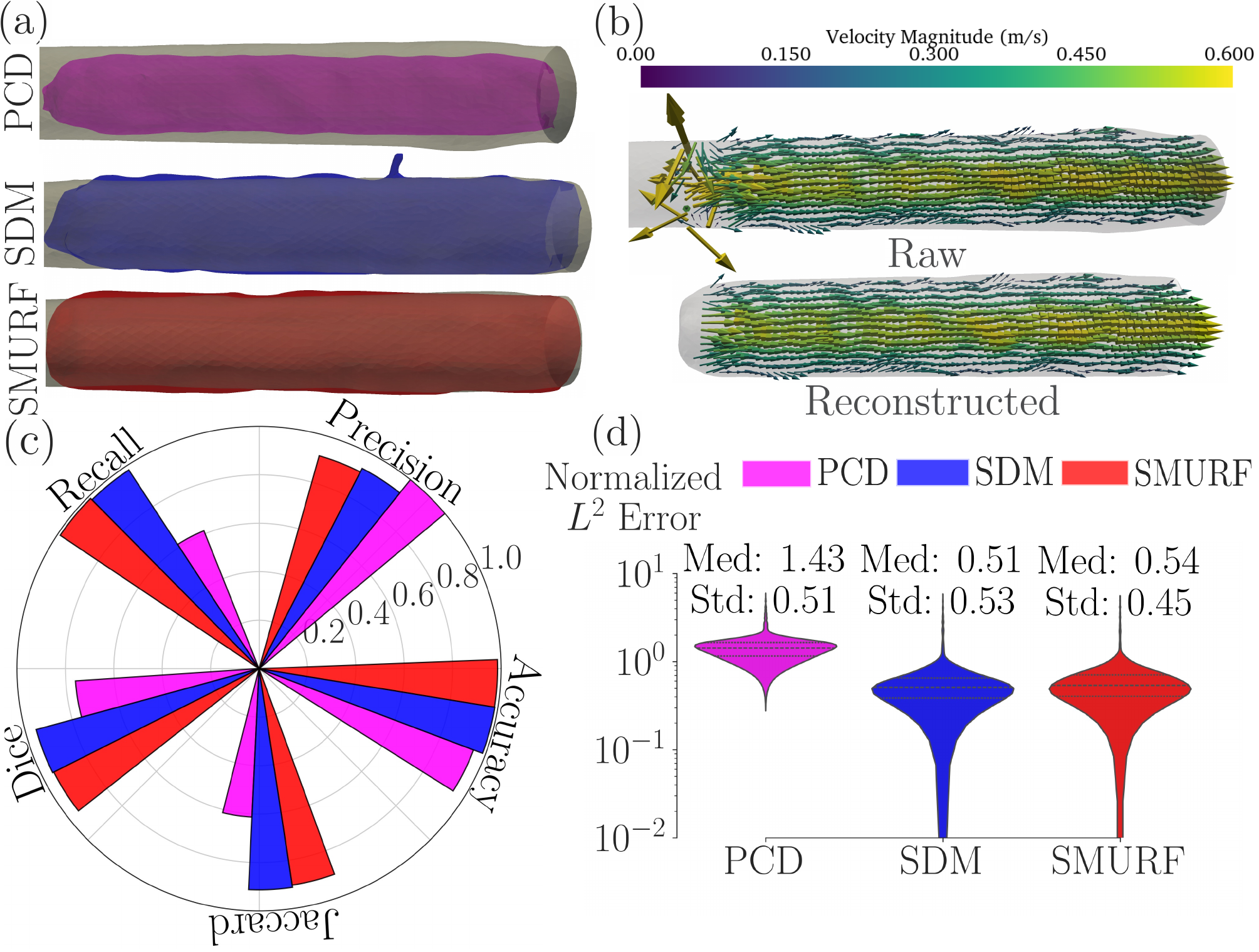}
        \caption{\textit{In vitro} Poiseuille flow experimental validation. 
        (a) Expert-segmented geometry overlaid with PCD, SDM, and SMURF segmentation results; 
        (b) Axial section velocity visualization comparing raw velocity measurements with expert segmentation (top row) versus SMURF-reconstructed velocities and geometry (bottom row); 
        (c) Evaluation of segmentation accuracy using volumetric classification scores (see Table~\ref{tab:seg_scores}); 
        (d) Normalized \(L^2\)-error violin plots quantifying deviation of predicted segmentations from expert annotation.} 
        \label{fig:invitro_fig_1}
\end{figure}

\textbf{Flow segmentation.}
We validate SMURF on \textit{in vitro} Poiseuille flow with a known velocity profile.
Fig.~\ref{fig:invitro_fig_1}a compares SMURF's segmentation performance with expert annotations, PCD, and SDM. 
PCD significantly underestimates the flow domain near the wall, deviating from its performance on synthetic data.
SDM segments the flow domain more accurately than PCD but remains sensitive to background noise, causing a wavy surface.
These surface irregularities complicate the evaluation of wall-based flow metrics such as WSS.
SMURF restricts segmentation to flow voxels, producing a surface that aligns more closely with the known geometry.

Fig.~\ref{fig:invitro_fig_1}(c–d) shows quantitative metrics that support these observations.
In Fig.~\ref{fig:invitro_fig_1}c, PCD appears to have inflated precision because its underpredicted region lies fully within the actual volume (see Table~\ref{tab:seg_scores}).
SMURF achieves the highest scores, followed by SDM, with most of their scores exceeding \(90\%\), demonstrating their effectiveness in classifying flow and background regions.

Fig.~\ref{fig:invitro_fig_1}d compares the normalized \(L^2\)-error between predicted segmentations and expert annotation, demonstrating that SMURF and SDM achieve similar low subvoxel median errors. 
While SMURF has a slightly higher median error than SDM, it exhibits a smaller error spread due to its significantly smoother surface. 
In contrast, PCD shows the largest deviations from expert annotations, with most wall points deviating by more than a voxel from the true surface.

\textbf{Flow reconstruction.}
We evaluate SMURF's velocity reconstruction using the axial section of the Poiseuille flow. 
Fig.~\ref{fig:invitro_fig_1}b shows the raw and SMURF-reconstructed velocities. 
Raw measurements include artifacts at the inlet and velocity vectors near the walls that violate the no-penetration boundary condition.
These artifacts reduce the accuracy of flow-derived metrics.
SMURF removes these spurious vectors and yields a field closer to the expected flow profile.

\begin{figure}
    \centering
    \includegraphics[width=1\linewidth]{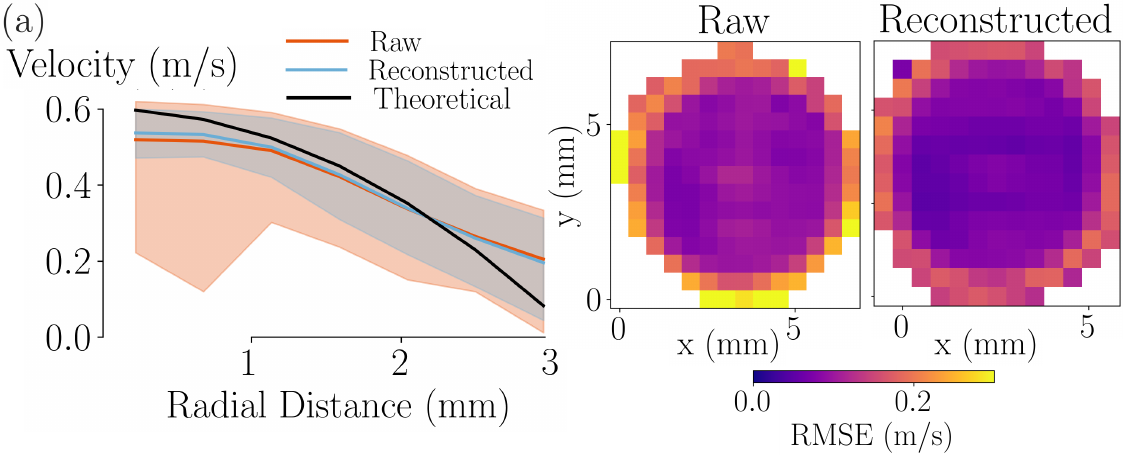}
    \caption{Quantitative evaluation of velocity reconstruction for \textit{in vitro} Poiseuille flow. 
    (a) Compares velocity magnitude surfaces to the theoretical paraboloid; 
    (b) Shows spatial velocity error distributions on a cross-sectional slice.}
    \label{fig:invitro_fig_4}
\end{figure}

Quantitative analyses in Fig.~\ref{fig:invitro_fig_4} highlight SMURF's performance. 
We bin voxel velocities radially from the centerline and plot them against the radial distance overlaid with the theoretical parabolic velocity profile in Fig.~\ref{fig:invitro_fig_4}a. 
Solid lines show mean velocities, while shaded regions indicate the 95th percentile velocity interval at each radius. 
Raw measurements deviate significantly, showing increased variance toward the core and elevated velocities near the wall. 
SMURF aligns the velocity profile more closely with theoretical predictions, reducing variance bands and lowering velocity near the wall, approaching the no-slip condition. 
A gap remains between SMURF’s mean velocity and the theoretical zero at the boundary, which we may address by enforcing physical constraints in future work.

Fig.~\ref{fig:invitro_fig_4}b presents spatial error distributions on a cross-sectional slice. SMURF produces a more uniform error distribution and reduces overall errors compared to raw measurements, particularly near the walls.

Table~\ref{tab:invitro_metrics_comparison} compares reconstruction metrics between raw and SMURF-reconstructed velocities. 
SMURF reduces RMSE by 33.9\%, improves structural similarity by 23.9\%, and enhances flow directionality by 5.9\%.
\begin{table}[ht]
    \centering
    \caption{Reconstruction metrics for raw and SMURF-reconstructed velocities compared to the theoretical profile for \textit{in vitro} Poiseuille flow.}
    \label{tab:invitro_metrics_comparison}
    \renewcommand{\arraystretch}{1.3} 
    \small
    \begin{tabular*}{\linewidth}{l@{\extracolsep{\fill}}ccc}
        \toprule
        \textbf{Metric} & \textbf{Raw} & \textbf{SMURF} & \textbf{Improvement (\%)} \\
        \midrule
        RMSE (m/s)          & 0.171 & 0.113 & 33.92  \\
        SSIM                & 0.327 & 0.405 & 23.85  \\
        Cosine Similarity   & 0.893 & 0.946 & 5.94   \\
        \bottomrule
    \end{tabular*}
\end{table}
%

\subsection{Demonstration on the in vivo internal carotid artery aneurysm}
\begin{figure}
    \centering
    \includegraphics[width=1\linewidth]{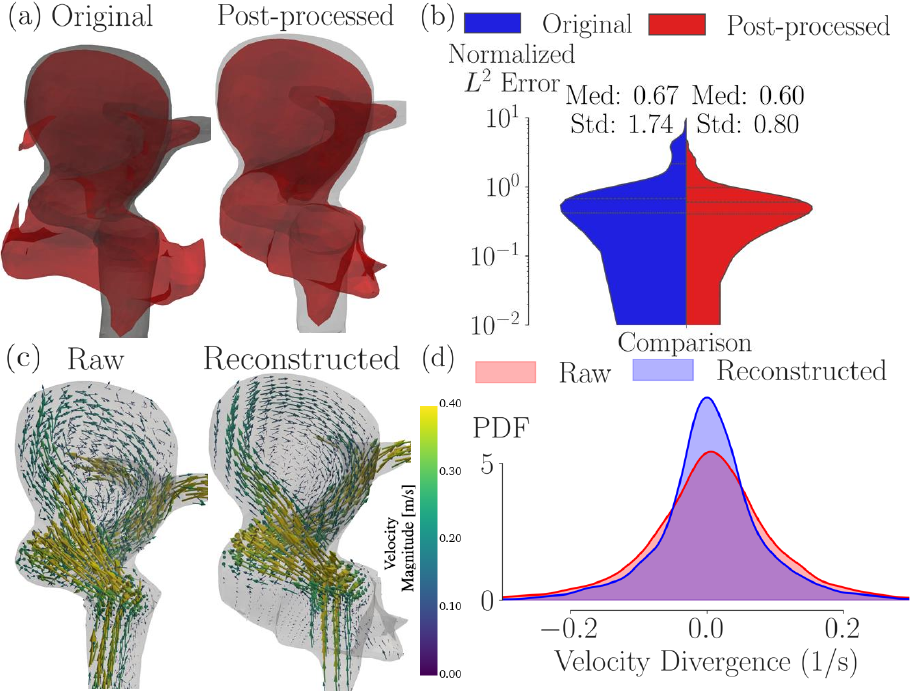}
    \caption{\textit{In vivo} ICA aneurysm segmentation and flow reconstruction using SMURF.
    (a) Original and post-processed SMURF segmentations versus expert TOF geometry, with post-processing used to isolate the ICA aneurysm region;
    (b) Normalized \(L^2\)-error comparison between original and post-processed SMURF segmentations relative to expert annotation; 
    (c) Peak systole velocity vector visualization comparing raw velocity data with expert-defined TOF geometry (left) and SMURF-reconstructed velocities with post-processed segmentation (right); 
    (d) Empirical probability density functions (PDFs) of velocity divergence residuals comparing raw and SMURF-reconstructed fields.}
    \label{fig:invivo_fig_1}
\end{figure}

\textbf{Flow segmentation.}  
We demonstrate SMURF's performance in an \textit{in vivo} setting using 4D flow data for a patient with an ICA aneurysm.
Fig.~\ref{fig:invivo_fig_1}a compares SMURF's segmentation with the expert-annotated TOF geometry.
SMURF captures the ICA and adjacent vessels due to similar flow and noise characteristics.

We post-process the segmentation to isolate the ICA region.
We first threshold the velocity magnitude at the 25th percentile to remove low-velocity voxels.
We then remove small connected components below the 99th percentile of connected component sizes and apply volumetric hole-filling.
These steps refine the segmentation, accurately isolating the ICA aneurysm region for further analysis.
We use \texttt{SciPy} and \texttt{scikit-image}~\cite{scipy,scikit-image} for post-processing.

SMURF effectively segments the ICA, including its anomalous features, such as the aneurysm dome. 
However, it under-segments the MCA and ACA after bifurcation, possibly because high velocities in smaller vessels cause misclassification (see Fig.~\ref{fig:invivo_fig_1}a).
Incorporating flow continuity and prior geometrical references may address this issue, which we will investigate in future work.

Table~\ref{tab:invivosegmentation_metrics} and Fig.~\ref{fig:invivo_fig_1}b show SMURF's quantitative performance.
Post-processing increases accuracy, precision, and composite metrics (Dice score and Jaccard index).
The original segmentation includes adjacent vessels, leading to lower precision, but post-processing achieves 98\% accuracy, 81\% precision, and a median \(L^2\)-error of 0.6 voxels, with 75\% of the surface within one voxel of expert annotations.
\begin{table}[ht]
    \centering
    \caption{Comparison of volumetric segmentation metrics (see Table~\ref{tab:seg_scores}) for SMURF's original and post-processed segmentations relative to expert annotations for the \textit{in vivo} ICA aneurysm dataset.}
    \label{tab:invivosegmentation_metrics}
    \renewcommand{\arraystretch}{1.3}
    \small
    \resizebox{\columnwidth}{!}{%
    \begin{tabular}{lccccc}
        \toprule
        \textbf{Segmentation} & \textbf{Accuracy} & \textbf{Precision} & \textbf{Recall} & \textbf{Dice score} & \textbf{Jaccard index} \\
        \midrule
        Original       & 0.94 & 0.44 & 0.77 & 0.56 & 0.39 \\
        Post-processed & 0.98 & 0.81 & 0.72 & 0.76 & 0.62 \\
        \bottomrule
    \end{tabular}%
    }
\end{table}

\textbf{Flow reconstruction.}  
Fig.~\ref{fig:invivo_fig_1}c compares raw and reconstructed velocity fields at peak systole. 
Raw measurements capture jets impinging on the aneurysm dome and swirling near the superior wall but contain noise that produces unphysical vectors in smaller flow regions.
SMURF denoises the field while preserving key flow features.

Fig.~\ref{fig:invivo_fig_1}d shows empirical PDFs of velocity divergence residuals estimated via kernel density estimation, which should ideally be zero for incompressible flow. 
SMURF reduces divergence values, drawing them closer to zero.
The median residual drops from \(0.073 \, \text{s}^{-1}\) (raw) to \(0.051 \, \text{s}^{-1}\) (reconstructed), a 30.6\% decrease, and the interquartile range shrinks by 27.1\%.
These findings indicate that SMURF mitigates noise-induced artifacts and improves physical consistency in velocity fields.

\section{Conclusion}\label{sec:conclusion}

We introduce SMURF, a scalable and unsupervised machine learning method for segmenting vascular geometries and reconstructing velocity fields from 4D flow MRI data. 
SMURF represents geometry and velocity fields with MLP-based functions that use Fourier feature embeddings and random weight factorization. 
A measurement model links the fields to the observed magnitude and velocity data. 
SMURF relies on maximum likelihood estimation and subsampling to handle large datasets.

SMURF has four key contributions. 
First, it uses MLP-based representations of fields, enabling super-resolution of segmented geometries and reconstructed velocity fields.
Second, it integrates geometry and velocity in a single measurement model, allowing simultaneous segmentation and reconstruction from magnitude data, velocity data, or both. 
Third, it employs subsampling to handle large, high-dimensional datasets without sacrificing computational efficiency. 
Finally, it operates in an unsupervised manner, removing the need for expert annotations and reducing potential sources of error.

We test SMURF on synthetic, \textit{in vitro}, and \textit{in vivo} data. 
In synthetic data, SMURF recovers velocities in high-noise conditions and maintains sub-voxel segmentation accuracy, outperforming or matching existing methods. 
In the \textit{in vitro} experiment, SMURF aligns reconstructed velocities with theoretical profiles, reduces velocity errors compared to raw measurements, and recovers underlying flow structures. 
In \textit{in vivo} data, SMURF isolates anatomical anomalies and decreases velocity divergence residuals, supporting the assessment of patient-specific 4D flow measurement.
Future work will extend flow validation by benchmarking against independent in vitro flow references, e.g., volumetric particle tracking velocimetry with stochastic tracking~\cite{hao2023unbalanced,hao2024unbalanced, singh2025new}.

In conclusion, the key contributions enable SMURF to advance data-driven diagnostics without expert labels, performing simultaneous segmentation and velocity reconstruction within a single framework while handling high-dimensional data. 
By doing this, SMURF provides a reliable approach to enhancing the diagnostic utility of 4D flow MRI in clinical settings.

\appendices
\section{Neural network architectures for geometry and velocity fields}\label{sec:neural_networks_archs}

We use separate neural networks configured with modified multilayer perceptrons (MLPs)~\cite{wang2023expert} to model the geometry and velocity field components. 
The forward pass in an \(L\)-layer modified MLP begins with two encoders for input coordinates: \( \mathbf{U} = \sigma(\mathbf{W}_1\mathbf{x} + \mathbf{b}_1), \quad \mathbf{V} = \sigma(\mathbf{W}_2\mathbf{x} + \mathbf{b}_2) \). After that, for \( l=1,2,\ldots, L \), we have \( \mathbf{f}^{(l)}(\mathbf{x}) = \mathbf{W}^{(l)} \cdot \mathbf{g}^{(l-1)}(\mathbf{x}) + \mathbf{b}^{(l)} \), followed by \( \mathbf{g}^{(l)}(\mathbf{x}) = \sigma(\mathbf{f}_{\theta}^{(l)}(\mathbf{x})) \odot \mathbf{U} + \bigl(1 - \sigma(\mathbf{f}_{\theta}^{(l)}(\mathbf{x}))\bigr) \odot \mathbf{V} \). Here, \(\sigma\) represents a nonlinear activation function, and \(\odot\) indicates element-wise multiplication. 
The final output of the network is given by \( \mathbf{f}_\theta(\mathbf{x}) = \mathbf{W}^{(L+1)} \cdot \mathbf{g}^{(L)}(\mathbf{x}) + \mathbf{b}^{(L+1)} \). 
The set of all trainable parameters consists of \( \{\mathbf{W}_1, \mathbf{b}_1, \mathbf{W}_2, \mathbf{b}_2, (\mathbf{W}^{(l)}, \mathbf{b}^{(l)})_{l=1}^{L+1}\} \).

We integrate Fourier feature embedding and Random Weight Factorization (RWF) into the modified MLP to learn complex functions essential for physical simulations. 
Fourier feature embedding mitigates spectral bias~\cite{rahaman2019spectral, xu2019frequency, basri2020frequency} and enhances the capture of high-frequency features, while RWF improves weight optimization in our neural networks.

The four-dimensional input vector, which encapsulates both spatial and temporal dimensions, passes through a Fourier feature embedding before entering the modified MLP. 
This technique transforms the input vectors into high-frequency signals before processing them through the modified MLP. 
The Fourier feature mapping function \( \gamma: \mathbb{R}^{d} \rightarrow \mathbb{R}^{2m} \) is formulated as \( \gamma(\mathbf{x}) = [\cos (\mathbf{B}\mathbf{x}), \sin(\mathbf{B}\mathbf{x})] \), where \( \mathbf{B} \in \mathbb{R}^{m \times d} \) is composed of entries drawn from a Gaussian distribution \( \mathcal{N}(0, \sigma^2) \) and \(\sigma>0\) is a user-specified hyperparameter. 
In our work, \( d=4 \) and we set the dimensionality of the high-frequency signal space to \( 2m = 256 \), comprising 128 sine and 128 cosine components. 
The standard deviation parameter is set to \(\sigma = 5.0\).

RWF decomposes the weights associated with each neuron as \( \mathbf{w}^{(k,l)} = s^{(k,l)} \cdot \mathbf{v}^{(k,l)} \), where \( k = 1,2,\ldots,d_l \) represents the index of neurons within layer \( l \) and \( l = 1,2,\ldots,L+1 \) denotes the layer index in a network with \( L \) layers. 
Here, \( \mathbf{w}^{(k,l)} \in \mathbb{R}^{d_{l-1}} \) denotes the weight vector representing the \( k \)-th row of the weight matrix \( \mathbf{W}^{(l)} \); the scalar \( s^{(k,l)} \in \mathbb{R} \) is a trainable scale factor for each neuron, and \( \mathbf{v}^{(k,l)} \in \mathbb{R}^{d_{l-1}} \) signifies the vector component of the weight. 
Consequently, this weight decomposition allows for expressing the weight matrix as \( \mathbf{W}^{(l)} = \text{diag}(\mathbf{s}^{(l)}) \cdot \mathbf{V}^{(l)} \) for \( l=1,2,\ldots,L+1 \), with \( \mathbf{s}^{(l)} \in \mathbb{R}^{d_l} \).

In practice, we apply RWF as follows. 
First, we initialize the parameters of the modified MLP using a conventional method like the Glorot scheme~\cite{glorot2010understanding}. Next, we create a scale vector \( \exp(\mathbf{s}) \) for each weight matrix \( \mathbf{W} \), where \(\mathbf{s}\) is sampled from a multivariate normal distribution \( \mathcal{N}(\mu, \sigma\mathbf{I}) \). 
Then, we factor each weight matrix by the corresponding scale factor as \( \mathbf{W} = \text{diag}(\exp(\mathbf{s})) \cdot \mathbf{V} \) during initialization. 
Finally, we optimize the new parameters \( \mathbf{s} \) and \( \mathbf{V} \). 
In our work, we set \(\mu = 0.5\) and \(\sigma = 0.1\). 
For more details on RWF, refer to~\cite{wang2022random,wang2023expert}.

Table~\ref{table:neural_net_arch} specifies the configurations for our neural network models.
\begin{table}[ht]
\centering
\caption{Neural network configurations for geometry ($g_{\psi}(r_i,t)$) and velocity components ($v_{\phi,1,2,3}(r_i,t)$).}
\label{table:neural_net_arch}
\setlength{\tabcolsep}{2pt}
\renewcommand{\arraystretch}{1.3}
\scriptsize
\begin{tabular*}{\columnwidth}{@{\extracolsep{\fill}}lccccc@{}}
\toprule
\textbf{Field} & \textbf{Input size} & \textbf{Output size} & \textbf{Depth} & \textbf{Width} & \textbf{Activation} \\
\midrule
\(g_{\psi}\) & 4 & \makecell{No.\\of classes} & 4 & 256 & Tanh \\
\(v_{\phi,1,2,3}\) & 4 & 1 & 4 & 256 & Tanh \\
\bottomrule
\end{tabular*}
\end{table}

\section*{Acknowledgment}
The authors thank Neal Minesh Patel for assisting with the \textit{in vitro} Poiseuille flow experiment and segmenting flow from time-of-flight images. 
They also thank Rudra Sethu Viji for preparing the PDMS block and the blood analog solution and supporting the \textit{in vitro} flow experiment.

The contributions of the authors are as follows:
\begin{itemize}
    \item Atharva Hans: Methodology, Software, Validation, Visualization, Writing – original draft.
    \item Abhishek Singh: Validation, Visualization, Writing – original draft.
    \item Pavlos Vlachos: Funding acquisition, Validation, Writing – review \& editing.
    \item Ilias Bilionis: Funding acquisition, Methodology, Validation, Writing – review \& editing.
\end{itemize}

\bibliographystyle{IEEEtran}
\bibliography{bibliography}

\end{document}